\documentclass[%
reprint,
superscriptaddress,
showpacs,preprintnumbers,
amsmath,amssymb,
aps,
prb,
floatfix,
notitlepage,
citeautoscript
]{revtex4-2}
\usepackage{array}
\usepackage{graphicx}
\usepackage[export]{adjustbox}
\usepackage{dcolumn}
\usepackage{bm}
\usepackage{amsmath} 
\usepackage{wasysym}
\usepackage[utf8]{inputenc}
\usepackage{color}
\usepackage{blindtext}
\usepackage{bm}
\usepackage{todonotes}
\usepackage{pgfplots}

\begin{document}
\newlength\fheight
\newlength\fwidth

\title{Excitonic magneto-optics in monolayer transition metal dichalcogenides: From nanoribbons to two-dimensional response}

\author{J. Have}
\email{jh@nano.aau.dk}
\affiliation{Department of Materials and Production, Aalborg University, DK-9220 Aalborg East, Denmark}
\affiliation{Department of Mathematical Sciences, Aalborg University, DK-9220 Aalborg East, Denmark}
\author{N.M.R. Peres}
\affiliation{International Iberian Nanotechnology Laboratory (INL), 4715-330 Braga, Portugal}
\affiliation{Center and Department of Physics, and QuantaLab, University of Minho, Campus de Gualtar, 4710-057 Braga, Portugal}
\author{T.G. Pedersen}
\affiliation{Department of Materials and Production, Aalborg University, DK-9220 Aalborg East, Denmark}
\affiliation{Center for Nanostructured Graphene (CNG), DK-9220 Aalborg East, Denmark}

\begin{abstract}
The magneto-optical response of monolayer transition metal dichalcogenides (TMDs), including excitonic effects, is studied using a nanoribbon geometry. We compute the diagonal optical conductivity and the Hall conductivity. Comparing the excitonic optical Hall conductivity to results obtained in the independent particle approximation, we find an increase in the amplitude corresponding to one order of magnitude when excitonic effects are included. The Hall conductivities are used to calculate Faraday rotation spectra for MoS\textsubscript{2} and WSe\textsubscript{2}. Finally, we have also calculated the diamagnetic shift of the exciton states of WSe\textsubscript{2} in different dielectric environments. Comparing the calculated diamagnetic shift to recent experimental measurements, we find a very good agreement between the two. 
\end{abstract}

\maketitle

\section{Introduction}
\label{sec:introduction}
With the successful exfoliation of monolayers of transition metal dichalcogenides \cite{mak2010atomically}, a new group of interesting semiconducting materials became available for study and potential applications. The characteristics of monolayer TMDs include a direct band gap \cite{mak2010atomically,splendiani2010emerging,zhang2014direct}, broken inversion symmetry \cite{xiao2012coupled,xu2014spin}, strong spin-orbit coupling \cite{liu2013three}, and strongly bound excitons and excitonic complexes \cite{berkelbach2013theory,ugeda2014giant,chaves2017excitonic,van2018excitons}. In addition to these characteristics, monolayer TMDs have also been shown to exhibit interesting magneto-optical properties such as valley polarized Landau levels \cite{wang2017valley,rose2013spin,chu2014valley}, valley Zeeman splitting \cite{li2014valley,macneill2015breaking,srivastava2015valley,wang2015magneto,aivazian2015magnetic}, and magnetic field induced rotation of the polarization state of light \cite{schmidt2016magnetic,wang2016control}. These properties have inspired potential new applications in areas such as optoelectronics \cite{sun2016optical} and valleytronics \cite{mai2013many,schaibley2016valleytronics}. Magnetic fields have also been used to probe exciton properties, such as effective mass, size \cite{stier2018magnetooptics,zipfel2018spatial,liu2019magneto}, and how they are affected by the dielectric environment \cite{stier2016probing}. 

So far, the theoretical analysis of TMD magnetoexcitons has relied on effective mass models, such as the Wannier model \cite{stier2016probing,stier2016exciton,stier2018magnetooptics,van2018excitons}. We recently validated that the Wannier model can be used to accurately describe certain properties of magnetoexcitons \cite{have2019monolayer}. However, in the Wannier model, the Bloch part of the wave function is replaced by a plane wave, which makes the task of computing the single-particle momentum matrix-elements unfeasible. For the diagonal optical response there is a solution to this problem \cite{pedersen2016exciton}, but for the Hall conductivity no solution currently exists. Thus, the Wannier model cannot be applied to the task of calculating the Hall conductivity, which is a necessary step in computing the magneto-optical Kerr effect and the Faraday rotation \cite{pedersen2003tight,morimoto2009optical}. The issue can be resolved in the independent particle approximation (IPA) \cite{pedersen2011tight,chu2014valley,catarina2019optical}, but the optical properties of TMDs are dominated by excitonic effects. Hence, for an accurate description of the magneto-optical response of TMDs, excitons should be included. 

The main computational difficulty in going beyond effective mass models when treating magnetoexcitons is that the external magnetic field breaks the translation symmetry of the system. Depending on the choice of magnetic vector potential gauge, translation symmetry will be broken in at least one direction. The translation symmetry can be restored by considering a magnetic supercell, but the size of the supercell is inversely proportional to the magnetic field strength\cite{pedersen2011tight}. Consequently, for realistic field strengths, a very large supercell is needed, thus, making the task of computing the excitonic properties unfeasible \cite{have2018magnetoexcitons}. In the present work, we address this issue by using a system of finite width in the direction, for which translation symmetry is broken. This approach corresponds to considering wide TMD nanoribbons. By increasing the size of the system in the finite direction, we are able to recover the two-dimensional (2D) response, including excitonic effects. Using this approach, we then describe quantitatively the excitonic effects on both the diagonal conductivity and the Hall conductivity of monolayer TMDs perturbed by an external magnetic field. This allows us to compute Faraday rotation spectra as well as excitonic diamagnetic shifts.

The paper is structured as follows: In Sec.~\ref{sec:ipa}, the tight-binding model used to describe the single-particle properties of both 2D monolayer TMDs and nanoribbons is introduced. In this section, we also check the width convergence of the nanoribbon optical response in the independent particle approximation. In Sec.~\ref{sec:exc-effects}, we include excitonic effects in our model and check convergence of the optical response. Finally, in Sec.~\ref{sec:results}, the magneto-optical response including excitons is studied using our nanoribbon model. In this section, we also calculate the diamagnetic shift of excitons in WSe\textsubscript{2} and compare to recent experimental results. 

\begin{figure}[t]
  \centering
  \includegraphics[width=0.48\textwidth]{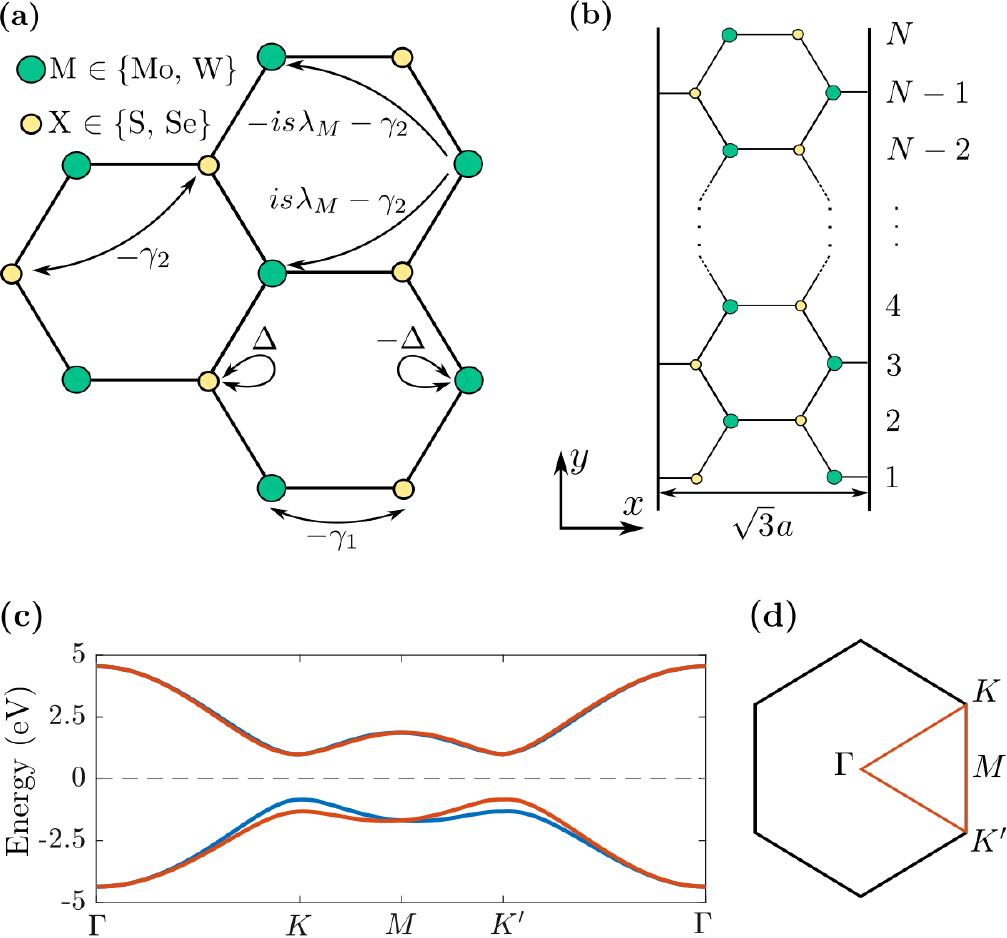}
  \caption{(a) Schematic of the tight-binding couplings in the NNN-TB model for monolayer MX\textsubscript{2} TMDs. (b) Unit cell of a MX\textsubscript{2} armchair nanoribbon of width $(N-1)a/2$ and length $\sqrt{3}a$, where $a$ is the lattice constant. (c) Band structure of WSe\textsubscript{2} along the path in the Brillouin zone specified by the letters. The blue and red lines are the spin-up and -down bands, respectively. (d) Brillouin zone of monolayer TMD.}\label{fig:couplings}
\end{figure}

\section{Single-Particle Properties}
\label{sec:ipa}
In this section, we present the theoretical framework used to describe the single-particle properties of monolayer TMDs and nanoribbons. Two important characteristics of monolayer TMDs are strong spin-orbit couplings (SOC) and broken electron-hole symmetry. These result in spin-splitting of the conduction and valence-band-edge states and different effective masses for electrons and holes \cite{xiao2012coupled,kormanyos2015k}. To include these characteristics in the description of the single-particle properties of monolayer TMDs, we apply a next-nearest-neighbor tight-binding (NNN-TB) model. Recently, a similar model was used in both the study of spin Hall effects in monolayer TMDs \cite{taghizadeh2018nonlinear} and to compute the optical response of gapped and proximitized graphene \cite{pedersen2018linear}. In addition to the nearest neighbor hopping term $\gamma_1$ used in Refs.~\onlinecite{pedersen2018linear,taghizadeh2018nonlinear}, we also include a finite spin-independent hopping term $\gamma_2$ between second-nearest neighbors. This additional hopping term is used to model the different effective masses of electrons and holes. The couplings for a TMD with lattice constant $a$ are illustrated in Fig.~\ref{fig:couplings}(a). Here, $-\Delta$ and $\Delta$ denote the on-site energies for transition metal ($M$) and chalcogen ($X$) atoms, respectively. Additionally, $is\lambda_M\eta$ is the spin-orbit coupling between second-nearest neighbor transition metal atoms \cite{kochan2017model}, where $s=\pm 1$ denotes the spin and $\eta=\pm 1$. The value of $\eta$ depends on the rotation sense in a hexagon, $\eta=+1(-1)$ for clockwise (counterclockwise) orientation. For simplicity, we assume that the spin-orbit coupling between chalcogen atoms is negligible. The same couplings are used to describe both TMD monolayers and nanoribbons.

\begin{table}[b]
  \centering
  \begin{tabular}{l|c c c c c c }
   & $\Delta$ (eV) & $\gamma_1$ (eV) & $\gamma_2$ (meV) & $\lambda_{M}$ (meV) & $a$ (\AA) & $r_0$ (\AA)  \\ \hline\hline
  MoS\textsubscript{2} & 1.24 & 1.498 & 8.2 & 14.4 & 3.18 & 44.3\\ \hline
  MoSe\textsubscript{2} & 1.09 & 1.359 & 92.5 & 18.3 & 3.32& 51.2\\ \hline
  WS\textsubscript{2} & 1.22 & 1.661 & -51.7 & 43.3 & 3.19 & 39.9\\ \hline
  WSe\textsubscript{2} & 1.04 & 1.444 & -43.6 & 48.5 & 3.32 & 46.2\\
\end{tabular}\caption{Model parameters for the four common TMDs. The on-site energy, lattice constants, and the screening lengths $r_0$ are taken from Ref.~\onlinecite{rasmussen2015computational}. The SOC strengths are calculated from the spin-splitting in Ref.~\onlinecite{rasmussen2015computational}, and the tight-binding couplings $\gamma_1$ and $\gamma_2$ are found by fitting to the electron and hole effective masses of Ref.~\onlinecite{rasmussen2015computational}.}
\label{tab:parameters}
\end{table}

We begin by considering a TMD monolayer placed in the $xy$-plane. The couplings described above give the following two-band Hamiltonian for a state with wave-vector $\mathbf{k}$:
\begin{equation}\label{eq:H0}
  \hat{H} = \left[
  \begin{array}{c c}
  \Delta -\gamma_2h & -\gamma_1 f\\
  -\gamma_1 f^{*} & -\Delta - s\lambda_Mg - \gamma_2h
  \end{array}  
  \right],
\end{equation}
where
\begin{align}
  f(\mathbf{k}) &= e^{ik_xa/\sqrt{3}}+2e^{-ik_xa/2\sqrt{3}}\cos(k_ya/2),\\
  g(\mathbf{k}) &= 2\left[\sin\left(\frac{k_x a\sqrt{3}}{2}+\frac{k_ya}{2}\right)-\sin(k_ya)\right.\nonumber\\
  &\qquad\left.-\sin\left(\frac{k_x a\sqrt{3}}{2}-\frac{k_ya}{2}\right)\right], \\
  h(\mathbf{k}) &= 2\left[\cos\left(\frac{k_x a\sqrt{3}}{2}+\frac{k_ya}{2}\right)+\cos(k_ya)\right.\nonumber\\
  &\qquad\left.+\cos\left(\frac{k_x a\sqrt{3}}{2}-\frac{k_ya}{2}\right)\right].
\end{align}
To determine the hopping parameters $\gamma_1$ and $\gamma_2$, we fit to the effective masses of electrons and holes in monolayer TMDs extracted from first-principles calculations in Ref.~\onlinecite{rasmussen2015computational}. When doing this, we assume for simplicity that $\lambda_M = 0$. Then, the energy bands are given by 
\begin{equation}\label{eq:bands}
  E_{\pm}(\mathbf{k}) = \gamma_2 h(\mathbf{k}) \pm \sqrt{\Delta^2+\gamma_1^2|f(\mathbf{k})|^2}.
\end{equation}
By expanding $E_{\pm}(\mathbf{k})$ around the $K$ point $(k_x,k_y)=2\pi(1/\sqrt{3}, 1/3)/a$ in the Brillouin zone (illustrated in Fig.~\ref{fig:couplings}(d)), we get the following relations between the hopping parameters and the effective masses:
\begin{align}
  \frac{3a^2\gamma_1^2}{8\Delta}+\frac{3a^2}{4}\gamma_2 &= \frac{\hbar^2}{2m_e^{*}},\label{eq:1}\\
  \frac{3a^2\gamma_1^2}{8\Delta}-\frac{3a^2}{4}\gamma_2 &= \frac{\hbar^2}{2m_h^{*}}.\label{eq:2}
\end{align}
Here, $m^{*}_{e(h)}$ is the effective electron (hole) mass and $\hbar$ is the reduced Planck's constant. Solving for $\gamma_1$ and $\gamma_2$ in Eqs.~\eqref{eq:1} and \eqref{eq:2} gives the hopping parameters. The spin-dependent band gaps at the $K$ and $K'$ points are given by $E_g = 2\Delta \mp 3\sqrt{3}s\lambda_M$, where $+$ ($-$) holds at the $K$ ($K'$) point. The value of the SOC parameter $\lambda_M$ is determined by matching the split to the spin-splitting of the valence-band-edge calculated in Ref.~\onlinecite{rasmussen2015computational}. The resulting band structure for WSe\textsubscript{2} is plotted in Fig.~\ref{fig:couplings}(c). In Table~\ref{tab:parameters}, we provide the complete set of parameters used for monolayer TMDs in the present paper. 

We introduce the external magnetic field by transforming the hopping integrals according to the Peierls substitution \cite{kohn1959theory}, which is simply the transformation $t \mapsto t_{ij} = t e^{i\phi_{ij}}$, where $t$ is equal to either $\gamma_1$, $\gamma_2$, or $\lambda_M$. The Peierls phase $\phi_{ij}$ is given by
\begin{equation}\label{eq:peierls-phase}
  \phi_{ij} = \frac{e}{\hbar}\int_{\mathbf{R}_i}^{\mathbf{R}_j}\mathbf{A}\cdot d\mathbf{l}.
\end{equation}
Here, $e$ is the elementary charge, $\mathbf{R}_i$ and $\mathbf{R}_j$ denote the location of atoms at site $i$ and $j$, respectively, and $\textbf{A}$ is the magnetic vector potential, related to the magnetic field by $\mathbf{B}=\nabla \times \textbf{A}$. We take the magnetic field to be given by $\mathbf{B}=B\hat{z}$, where $B$ is the magnetic field strength. For 2D systems the phase factor evidently breaks translation symmetry, but it can be restored by using a suitable magnetic supercell \cite{pedersen2013hofstadter}. As mentioned in Sec.~\ref{sec:introduction}, the relation between field strength and the supercell size makes the calculation of excitonic properties unfeasible for realistic magnetic fields. However, for a nanoribbon system, which is finite in the $y$-direction, the Landau gauge, $\mathbf{A}=-By\hat{x}$, does not affect the translation symmetry of the system \cite{have2018magnetoexcitons}. Hence, no restrictions on the magnetic field strength and no magnetic supercell are required. This is the motivation for using nanoribbons as a tool to describe the magneto-optical response of monolayer TMDs for arbitrary magnetic field strengths. We will consider armchair nanoribbons, which are infinite in the $x$-direction and have a finite width of $W = (N-1)a/2$, where $N$ is the number of dimer lines in the $y$-direction. The unit cell is illustrated in Fig.~\ref{fig:couplings}(b). By increasing $N$, we expect the optical response of the nanoribbons to converge to that of the 2D system.

\begin{figure}[t]
  \centering
  \includegraphics[width=0.49\textwidth]{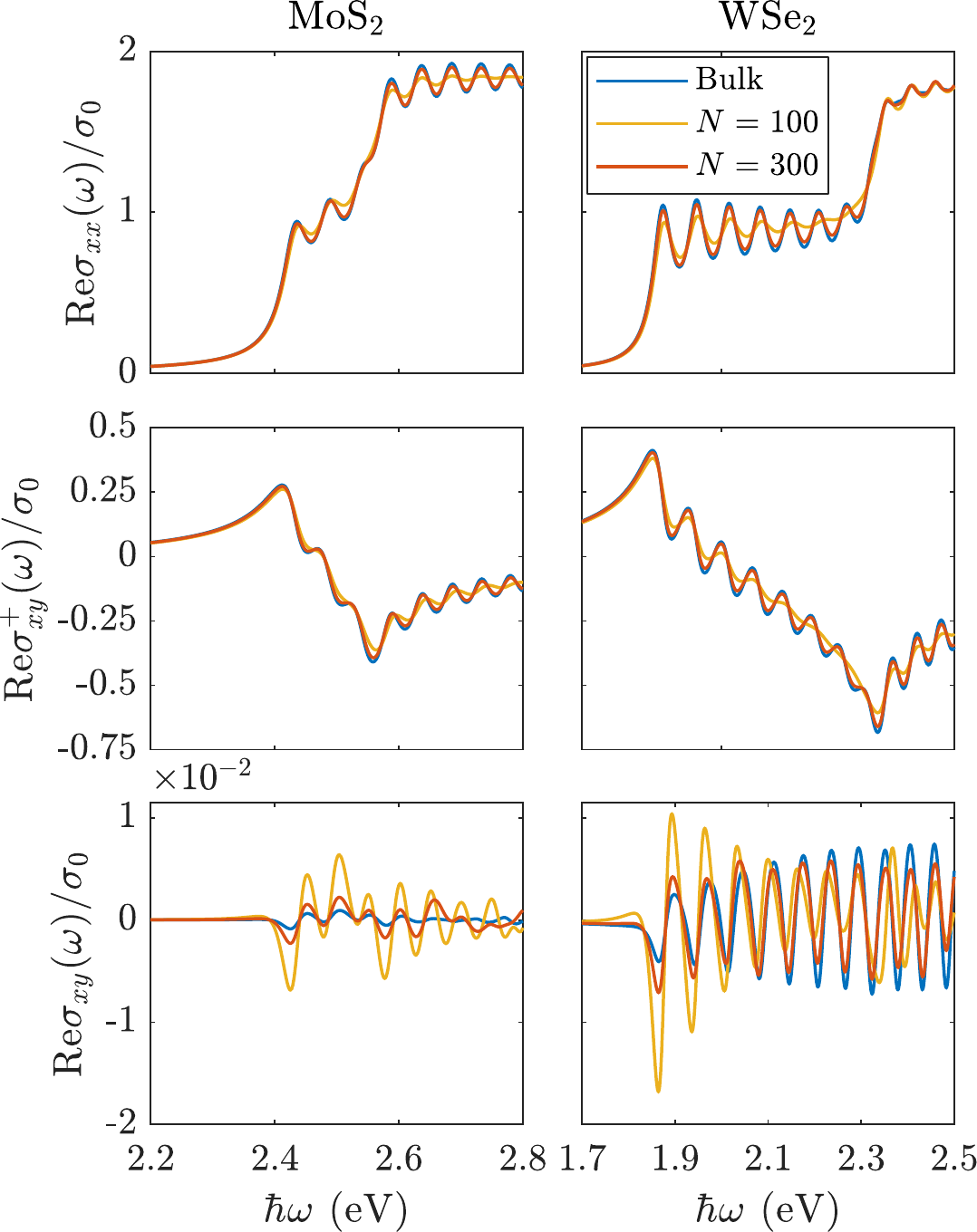}
  \caption{Single-particle linear optical conductivities of MoS\textsubscript{2} and WSe\textsubscript{2} for $B=130$ T and $\hbar\Gamma = 25$ meV. The blue lines refer to the 2D conductivities and red and yellow lines to the nanoribbon case.}\label{fig:converge-ipa}
\end{figure}

To calculate the linear optical conductivity, we make use of the following expression for the spin-up and -down contribution to the linear optical conductivity \cite{pedersen2011tight}
\begin{equation}\label{eq:cond}
  \sigma^{s}_{\alpha\beta}(\omega) = -\frac{ie^2\hbar^2\omega}{m^2A}\sum_{cvk}\frac{p^\alpha_{cvk,s}p^{\beta}_{vck,s}}{E_{cvk,s}^2(E_{cvk,s}-\hbar\omega-i\hbar\Gamma)},
\end{equation}
with $\alpha,\beta \in \{x,y\}$, $\hbar\omega$ the photon energy, $m$ the free electron mass, $A = WL$ the system area, where $W$ is the system width and $L$ is the system length, $p^\alpha_{cvk,s}$ the momentum matrix elements, and $E_{cvk,s} := E_c(k,s)-E_v(k,s)$ the transition energy. The sum runs over all combinations of conduction ($c$) and valence ($v$) bands and $k$-points, and we have neglected the non-resonant term of the conductivity. In the nanoribbon geometry, the limit where $L$ goes to infinity should be taken. In practice this is done by converting the sum over $k$-points to an integral by using that the distance between two $k$-points is equal to $\Delta k = 2\pi/L$. The linear optical conductivity tensor elements are then found by summing over spin, i.e. $\sigma_{\alpha\beta}(\omega) = \sigma^{+}_{\alpha\beta}(\omega)+\sigma^{-}_{\alpha\beta}(\omega)$. By symmetry, we have the relation $\sigma^{+}_{\alpha\alpha}(\omega) = \sigma^{-}_{\alpha\alpha}(\omega)$ for the diagonal elements, and $\sigma^{+}_{\alpha\beta}(\omega) = -\sigma^{-}_{\alpha\beta}(\omega)$ for the off-diagonal elements when $B=0$~T. We note that the expression in Eq.~\eqref{eq:cond} holds for both nanoribbons and 2D monolayers, but $k$ denotes a scalar quantity in the former case and a vector quantity in the latter case. 

In Fig.~\ref{fig:converge-ipa}, we show the real part of the optical conductivity in the single-particle approximation for 2D monolayers and nanoribbons. All spectra are  plotted in units of $\sigma_0 = e^2/4\hbar$ and calculated for a Brillouin zone discretized using $120$ $k$-points. Throughout, we focus on MoS\textsubscript{2} and WSe\textsubscript{2} as examples of monolayer TMDs, but similar results hold for other types of TMDs. Landau levels (LL) are clearly visible in both the diagonal and off-diagonal response. The plots also illustrate the finite off-diagonal conductivities, the so-called Hall conductivities, present when there is an external magnetic field. Comparing the response of the $N=100$ and $N=300$ nanoribbons to the bulk conductivity, we see that for $\sigma_{xx}(\omega)$ and $\sigma_{xy}^{+}(\omega)$ both nanoribbon widths capture the qualitative features. However, the wider nanoribbons more accurately capture the position of the higher Landau levels and the amplitudes of the peaks. In contrast, for $\sigma_{xy}(\omega)$ very wide nanoribbons are needed to obtain good convergence of the amplitudes. This is due to the fact that $\sigma_{xy}(\omega)$ is the sum of $\sigma_{xy}^{+}(\omega)$ and $\sigma_{xy}^{-}(\omega)$, both of which are much bigger in amplitude than $\sigma_{xy}(\omega)$. Thus, while the difference between the spin-dependent off-diagonal response of nanoribbons and $2D$ is small compared to the amplitude of $\sigma^{+}_{xy}(\omega)$ it is large compared to the amplitude of the Hall conductivity. Consequently, making the Hall conductivity susceptible to poor convergence. The excitonic spectra are expected to show better convergence since the optical response is dominated by excitons, and the excitons are strongly localized in TMDs \cite{berkelbach2013theory,ugeda2014giant,he2014tightly}. Finally, we note that the valley Zeeman splitting is not described by the TB Hamiltonian in this paper.

\section{Excitonic effects}
\label{sec:exc-effects}
In this section, we include excitonic effects in our description of TMD monolayers and nanoribbons. The approach follows that of Refs.~\onlinecite{have2018magnetoexcitons,trolle2014theory}. We expand the excitonic wave function $|exc\rangle$ in a basis of singlets formed by excitations between a single pair of spin-dependent valence and conduction bands at $k$, such that the wave function is given by
\begin{equation}\label{eq:wavefunction}
|exc\rangle = \sum_{cvk,s}A^s_{cvk}|vks \to cks\rangle,
\end{equation}
where $A^s_{cvk}$ are the expansion coefficients and $|vks \to cks\rangle$ the singly excited states. Note, that we only include excitations between bands of equal spin and that $k$ can be either a vector or scalar quantity depending on the dimensionality of the system under consideration. The excitonic states are governed by the Bethe-Salpeter equation (BSE) \cite{berkelbach2013theory}, which for the expansion in Eq.~\eqref{eq:wavefunction} take the following form
\begin{equation}
 E_{cvk,s}A^s_{cvk} + \sum_{c'v'k',s'}W^{s,s'}_{cvk,c'v'k'}A^{s'}_{c'v'k'} = E A^s_{cvk}.\label{eq:BSE}
\end{equation}
Here, $W^{s,s'}_{cvk, c'v'k'}$ is the electron-hole interaction matrix-elements and $E$ is the exciton energy. Note, that we have neglected the exchange term in the BSE for simplicity. Then, the electron-hole interaction matrix-elements are given by
\begin{equation}\label{eq:eh-interaction}
  W^{s,s'}_{cvk,c'v'k'} = \langle vks \to cks |U| v'k's' \to c'k's'\rangle,
\end{equation}
where $U$ is the electron-hole interaction potential defined below. Performing the spin-integral in Eq.~\eqref{eq:eh-interaction}, we find
\begin{align}
  W^{s,s'}_{cvk,c'v'k'} = &\delta_{s,s'}\iint\mathrm{d}^2\mathbf{r}\mathrm{d}^2\mathbf{r}'\phi_{cks}^{*}(\mathbf{r})\phi_{vks}(\mathbf{r}') \nonumber\\
  &\times U(\mathbf{r}-\mathbf{r'})\phi_{c'k's}(\mathbf{r})\phi^{*}_{v'k's}(\mathbf{r}').\label{eq:eh-mel}
\end{align}
Here, $\phi_{\alpha ks}(\mathbf{r})$ are the tight-binding states with $\alpha \in \{c,v\}$. Equation \eqref{eq:eh-mel} shows that the spin-up and -down equations decouple and can be solved independently. 

In a strict 2D system the electron-hole interaction is not the usual Coulomb potential, but instead modeled by the Keldysh potential \cite{keldysh1979coulomb, trolle2017model}
\begin{equation}
  U(\mathbf{r}) = -\frac{e^2}{8\varepsilon_0r_0}\left[H_0\left(\frac{\kappa r}{r_0}\right)-Y_0\left(\frac{\kappa r}{r_0}\right)\right].
\end{equation}
Here, $\varepsilon_0$ is the vacuum permittivity, $H_0$ and $Y_0$ are Struve and Neumann functions, respectively, $r=|\mathbf{r}|$, $r_0$ is an in-plane screening length, and $\kappa$ is the average of the relative dielectric constant of the substrate and capping material. The values of $r_0$ used in this paper are listed in Tab.~\ref{tab:parameters}. For the strict 2D system, a straightforward calculation of the matrix-elements in Eq.~\eqref{eq:eh-mel} can be done using the approach of Ref.~\cite{trolle2014theory}. For the nanoribbon geometry, additional considerations are needed. We want the excitonic properties in the nanoribbon geometry to converge to those of the 2D system, when the ribbon width is sufficiently large. Thus, we need to modify the approach of Ref.~\cite{trolle2014theory} to work for structures, which are periodic in one direction, but have non-negligible width. The details are provided in Appendix~\ref{app:commutators}, but the main result is that for the nanoribbon geometry the matrix-elements $W^{s,s}_{cvk,c'v'k'}$ can be computed from
\begin{equation}
  W^{s,s}_{cvk,c'v'k'} = \sum_{n,m}I^{n,s}_{c k, c' k'}I^{m,s}_{v' k', v k}U^{k,k'}_{n,m}.
\end{equation}
Here $n$ and $m$ run over the atomic sites in the unit cell and $I^{n,s}_{\alpha k, \beta k'} = C^{n*}_{\alpha ks}C^{n}_{\beta k's}$ is the Bloch overlap given by the product of the tight-binding eigenvector elements belonging to site $n$. Finally, the integral factor $U^{k,k'}_{n,m}$ is defined as
\begin{equation}\label{eq:U_factor}
  U^{k,k'}_{n,m} = -\frac{e^2}{2\pi L \varepsilon_0}\int_0^\infty\mathrm{d}z K_0\left(\sqrt{r_0^2z^2+Y_{nm}^2}\left|k-k'\right|\right)e^{-\kappa z},
\end{equation}
where $K_0$ is a modified Bessel function of the second kind and $Y_{nm}=Y_n-Y_m$ is the difference between the $y$-coordinates of the atoms belonging to orbitals $n$ and $m$. The integral in Eq.~\eqref{eq:U_factor} is computed numerically using a suitable Gauss-quadrature.

The eigenvalue problem defined in Eq.~\eqref{eq:BSE} can be solved by diagonalization. Due to the decoupling of spin-up and -down equations, the matrix to be diagonalized is block diagonal. Thus, to obtain the full solution two eigenvalue problems of dimension $N_c N_v N_k$ have to be solved. Here, $N_c$ and $N_v$ are the number of conduction and valence bands, respectively, and $N_k$ is the number of $k$-points. For a magnetic field of $100$ T the 2D magnetic supercell consists of roughly 2000 atoms, hence making diagonalization of the BSE problem computationally unfeasible. On the other hand, using nanoribbons as a theoretical tool the linear optical response converges to the bulk 2D response when the nanoribbon unit cell contains roughly 200 atoms. The result is that the computations are feasible, although still very demanding.
However, if only the optical response and not the full eigenvalue decomposition is needed a significant reduction in computational complexity can be obtained by using the Lanczos approach in Refs.~\onlinecite{trolle2014theory,have2018magnetoexcitons}.

\begin{figure}[t]
  \centering
  \includegraphics[width=0.47\textwidth]{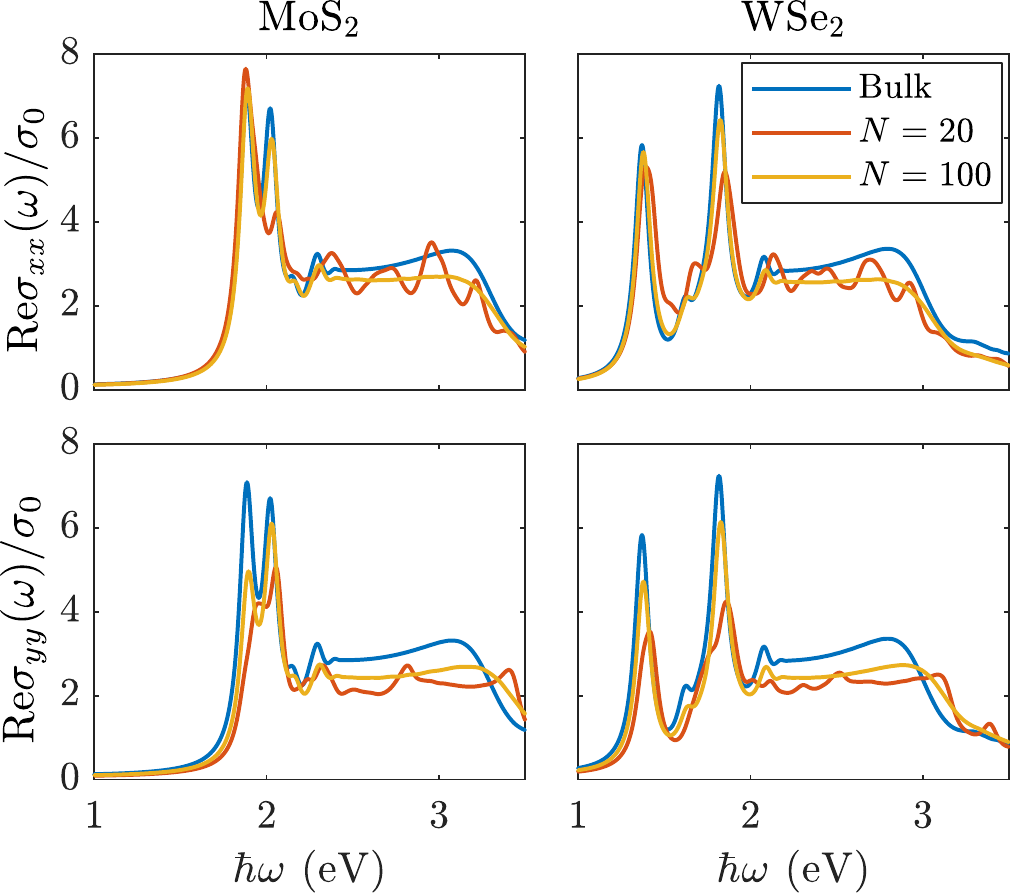}
  \caption{Many-body diagonal conductivity of MoS\textsubscript{2} and WSe\textsubscript{2} for $B=0$ T, $\hbar\Gamma = 50$ meV, and $\kappa = 1$. The blue lines refer to the 2D conductivities and yellow and red lines to the nanoribbon spectra.}\label{fig:exc-convergence}
\end{figure}

The Lanczos routine is based on the fact that real part of the linear optical conductivity can be computed from the expression \cite{trolle2014theory}
\begin{equation}\label{eq:exc-cond}
  \mathrm{Re}\sigma_{\alpha \beta}(\hbar\omega) = -\frac{e^2}{m^2\omega A}\sum_{s} \mathrm{Im}\langle P_{\alpha s} |\hat{G}_{s}(\hbar\omega)|P_{\beta s}\rangle,
\end{equation}
with $\alpha,\beta\in\{x,y\}$, $\hat{G}_{s}(\hbar\omega)$ the many-body Green's function given below, and $P_{\alpha s}$ given by
\begin{equation}
  \left|P_{\alpha s}\right\rangle := \hat{P}_\alpha |0,s\rangle = \sqrt{2}\sum_{cvk}A^s_{cvk}p^\alpha_{cvk,s}.
\end{equation}
Here, $|0,s\rangle$ is the many-body ground state, $\hat{P}_\alpha$ is the many-body momentum operator, and $p^\alpha_{cvk,s}$ denote the single-particle momentum matrix elements. The many-body Green's function in Eq.~\eqref{eq:exc-cond} is given by
\begin{equation}
  \hat{G}_{s}(\hbar\omega) = \lim_{\hbar\Gamma \to 0^+}(\hbar\omega+i\hbar\Gamma - \hat{H}_{s})^{-1},
\end{equation}
where $\hat{H}_{s}$ is the many-body Hamiltonian. In practice, we allow a small finite $\hbar\Gamma$ to add broadening to the spectra. The matrix elements of the Green's function in Eq.~\eqref{eq:exc-cond} are evaluated effectively as in Ref.~\onlinecite{have2018magnetoexcitons}, i.e. using the Lanczos-Haydock routine for tridiagonalization \cite{haydock1980recursive}. Computationally this is still a daunting task due to the size of the problem. For a nanoribbon with $N=100$ and using a discretization with $N_k = 120$, the matrix that is to be tridiagonalized has dimension $1.2\cdot10^6 \times 1.2\cdot10^6$. We reduce the size of the problem by disregarding the top and bottom half of the conduction and valence bands, respectively, which primarily affect the high-energy part of the spectra.

\begin{figure}[t]
  \centering
  \includegraphics[width=0.48\textwidth]{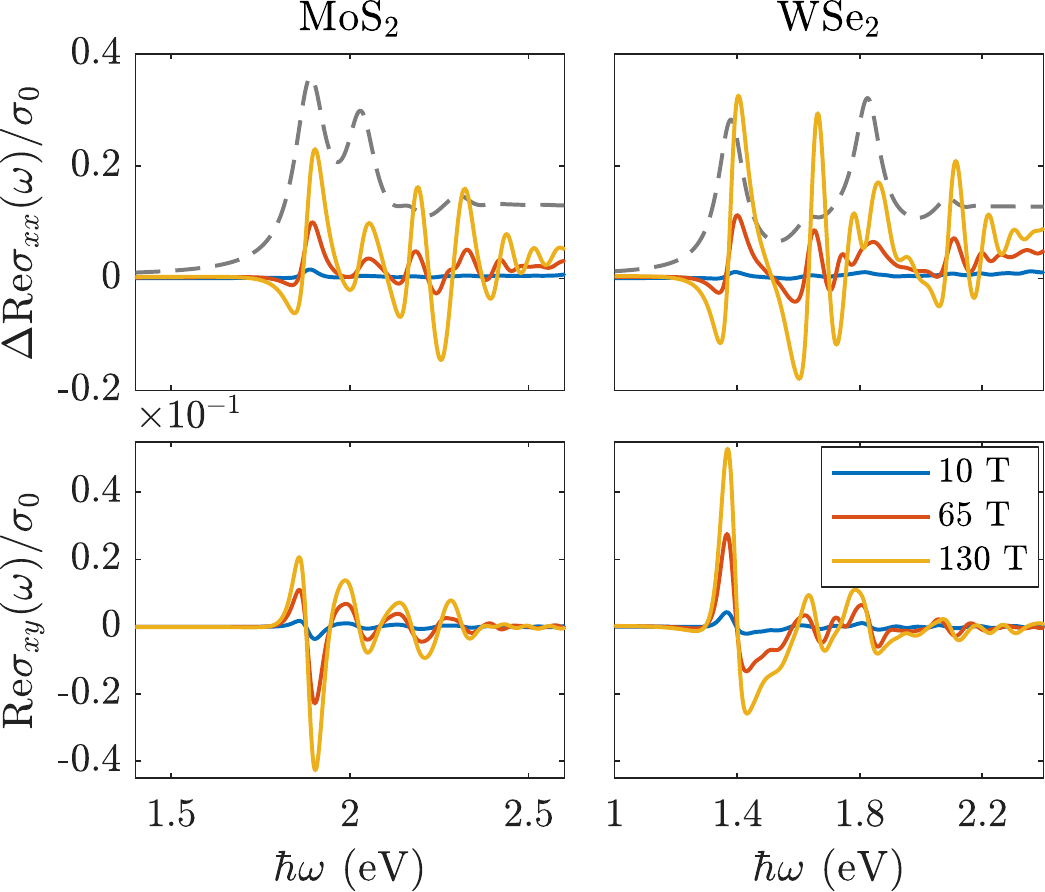}
  \caption{Excitonic optical conductivity of MoS\textsubscript{2} and WSe\textsubscript{2} as function of the external magnetic field. The first row illustrates $\Delta \mathrm{Re}\sigma_{xx}(\omega)$, which is the difference between the diagonal conductivity at a finite magnetic field strength and at 0 T. The dashed grey lines show the unperturbed spectra. The second row are the Hall conductivities at different magnetic field strengths. Spectra are for nanoribbons with $N=100$ and $\kappa = 1$.}\label{fig:exc-mag}
\end{figure}

In Fig.~\ref{fig:exc-convergence}, we show the convergence of the nanoribbon conductivities to the 2D response in the unperturbed case ($B=0$). The two main exciton peaks at $1.88$ eV and $2.02$ eV for MoS\textsubscript{2} and at $1.37$ eV and $1.82$ eV for WSe\textsubscript{2} are denoted by $A$ and $B$, respectively. The results show a good convergence for the nanoribbon with $N=100$. Both the $A$ and $B$ exciton peaks coincide with the bulk results and the peaks corresponding to the excited states also match the bulk results. The discrepancy at high photon energies are due to us disregarding some bands in the excitonic calculations. Regarding the amplitude of the peaks, we see that the amplitude is close to the bulk result for $\mathrm{Re}\sigma_{xx}$, while the $\mathrm{Re}\sigma_{yy}$ results could be improved by using wider nanoribbons. However, as our goal is to study the effect of an external magnetic field on the optical response, the convergence shown in Fig.~\ref{fig:exc-convergence} is satisfactory. Comparing to the spectra of unperturbed TMDs in Ref.~\onlinecite{chaves2017excitonic}, we see that the qualitative features agree well.

\section{Results}
\label{sec:results}
In this section, we present the results obtained from the theoretical framework of Secs.~\ref{sec:ipa} and \ref{sec:exc-effects}. All calculations are based on the nanoribbon geometry and are calculated for nanoribbons with $N=100$. This  corresponds to a ribbon width of $15.7$ nm and $16.4$ nm for MoS\textsubscript{2} and WSe\textsubscript{2}, respectively. The one-dimensional Brillouin zone is discretized using $120$ $k$-points. For all spectra, a broadening of $\hbar\Gamma = 50$ meV is used.


In Fig.~\ref{fig:exc-mag}, the first row of plots show the change of the real part of the diagonal conductivity as a function of the magnetic field strength relative to the zero field case. To illustrate this, we have plotted the difference between the diagonal conductivity at a finite magnetic field strength and at 0 T. The plots show that the exciton peaks in MoS\textsubscript{2} and WSe\textsubscript{2} exhibit a small blueshift in response to the applied magnetic field. This small but important phenomena is what allows for experimental estimation of the spatial extent and effective mass of excitons. We will evaluate the size of the shift and discuss this in detail below. In addition to the blueshift of the peaks, the amplitudes also increase slightly as the field strength increases. Comparing to the amplitude of the peaks in the unperturbed spectra in Fig.~\ref{fig:exc-convergence}, the increase in amplitude due to the magnetic field is only a few percent for a field strength of 130 T. Thus, both effects are small changes to the unperturbed results. Finally, in the high energy part of the spectra, the results show the emergence of an oscillating modulation appearing at strong magnetic fields. These oscillations correspond to transitions between Landau levels.

\begin{figure}[t]
  \centering
  \includegraphics[width=0.47\textwidth]{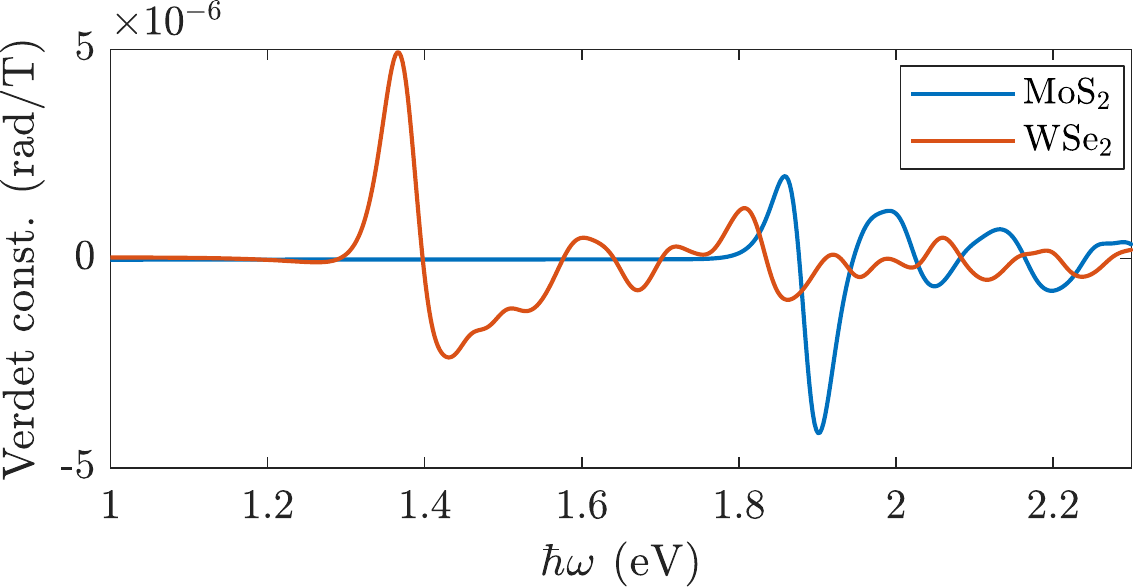}
  \caption{Plot of the Verdet constant for normal incident light on a TMD monolayer in vacuum.}\label{fig:verdet-const}
\end{figure}

The second row of plots in Fig.~\ref{fig:exc-mag} shows the Hall conductivities of MoS\textsubscript{2} and WSe\textsubscript{2}. In contrast to the diagonal conductivity, the magnetic field significantly alters the off-diagonal conductivities. Just as in the single-particle case, the time-reversal symmetry present in the absence of an external magnetic field ensures that the Hall conductivities are identical zero. When time-reversal symmetry is broken by the external magnetic field finite Hall conductivities are found even at small magnetic field strengths. This is illustrated in Fig.~\ref{fig:exc-mag} for MoS\textsubscript{2} and WSe\textsubscript{2}. Comparing the excitonic magneto-optical response in Fig.~\ref{fig:exc-mag} to the IPA results in Fig.~\ref{fig:converge-ipa}, we see that excitonic effects change the optical response significantly. In addition to changing the overall shape of the spectra, we also see that the excitonic Hall conductivities are approximately one order of magnitude larger than the IPA response. Hence, for an accurate description of the magneto-optical properties of monolayer TMDs, it is clearly important to account for excitons. Regarding the magnetic field dependence of the Hall conductivities in Fig.~\ref{fig:exc-mag}, we see that the amplitude scales linearly with the magnetic field strength. However, as we go to stronger fields, small changes in the shape of the spectra occur. These changes are due to the emergence of Landau levels and additional effects that are non-linear in $B$, such as the diamagnetic shift. 

\begin{figure}[t]
  \centering
  \includegraphics[width=0.48\textwidth]{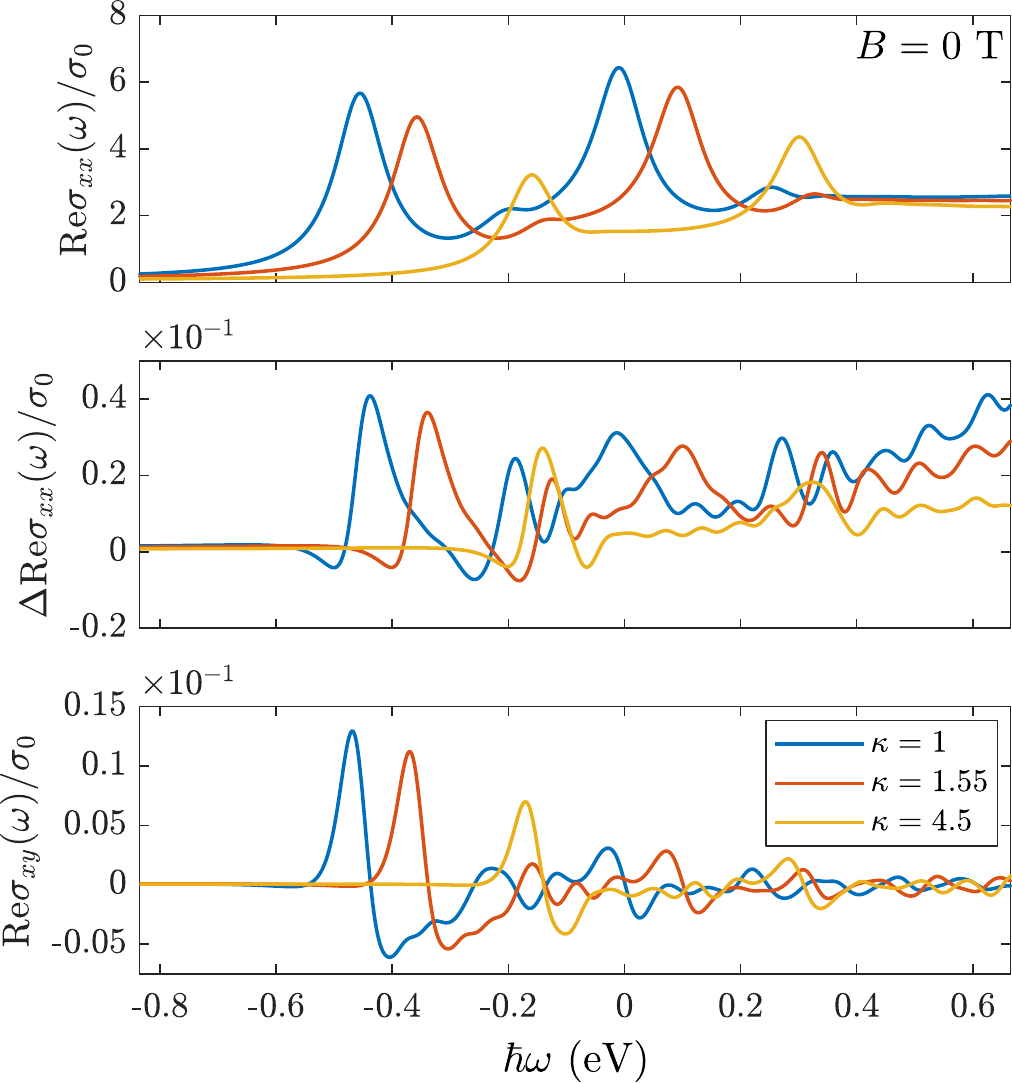}
  \caption{Excitonic optical conductivity of WSe\textsubscript{2} in different dielectric environments. The top panel shows the diagonal conductivity in the unperturbed case. The middle panel shows the change in the diagonal conductivity from the unperturbed case to the $B=30$ T case. The bottom panel shows the Hall conductivities calculated at $B=30$ T.}\label{fig:dielectric-mag}
\end{figure}

The finite Hall conductivity, present when there is an external magnetic field, causes the system to exhibit a magneto-optical Kerr effect (MOKE) and a Faraday effect. The MOKE is a rotation of the polarization state of light when reflected off the surface of a magnetized material, while the Faraday effect is a rotation of the polarization of the transmitted light. Here, we compute the Faraday rotation angle $\theta$ for normal incidence of light on a single layer of TMD. The rotation angle for a single passage of the monolayer can be approximated by \cite{morimoto2009optical,ferreira2011faraday}
\begin{equation}\label{eq:faraday}
\theta = \frac{1}{(n_1+n_2)c\varepsilon_0}\mathrm{Re}\sigma_{xy}(\omega),
\end{equation}
where $n_1$ and $n_2$ are the refractive index of the substrate and capping material, respectively, and $c$ is the speed of light. The expression in Eq.~\eqref{eq:faraday} is valid when $\sigma_{xx} \gg \sigma_{xy}$. As the Hall conductivity scales linearly with $B$ at small field strengths, the Faraday rotation angle is often expressed as $\theta = V B$, where $V$ is the so-called Verdet constant. In Fig.~\ref{fig:verdet-const}, we have computed the Verdet constant for freestanding MoS\textsubscript{2} and WSe\textsubscript{2}. As shown by the figure, the rotation for a single passage of the TMD monolayer is very small. However, this could be increased by placing the monolayer in an optical cavity in order to enhance the rotation by multiple passes \cite{ferreira2011faraday,da2018cavity}.


As mentioned in Sec.~\ref{sec:exc-effects}, the electron-hole interaction is screened by the substrate and capping materials. This screening is described by the $\kappa$ parameter, which is simply the average of the relative dielectric constant of the substrate and capping material. In Fig.~\ref{fig:dielectric-mag}, the optical conductivity of WSe\textsubscript{2} is shown for $\kappa$ values of 1, 1.55, and 4.5. These values correspond to WSe\textsubscript{2} placed  
in vacuum, on a SiO\textsubscript{2} substrate, or encapsulated in hBN, respectively. 
It should be noted, that an exchange self-energy correction to the single-particle band gap exists, and this effect is not included in our simple model. The self-energy correction is decreases when the screening from the surroundings increases \cite{catarina2019optical,have2019monolayer}. To account for this missing effect, the spectra in Fig.~\ref{fig:dielectric-mag} are shifted by the band gap energy. This allows us to observe changes in exciton binding energy as a function of $\kappa$. The first plot is of the diagonal conductivity for $B=0$ T, and the results show a blueshift of the exciton peaks as $\kappa$ increases. This is due to a decrease in the exciton binding energy as the screening from the surroundings is increased. The binding energy decreases from $455$ meV to $160$ meV as $\kappa$ increases from $1$ to $4.5$. The second row shows the change in the diagonal conductivity between the unperturbed case and the $B=30$ T case. The plots show that the diamagnetic shift of the $2s$ exciton states becomes harder to observe at higher values of $\kappa$, and that the Landau levels are not affected by the dielectric environment. Finally, the last plot is of the Hall conductivities. Here, the same blueshift is observed as in the diagonal conductivity. When going to the limit $\kappa \to \infty$, we recover the IPA results, as has been checked numerically. 

\begin{figure}[t]
  \centering
  \includegraphics[width=0.48\textwidth]{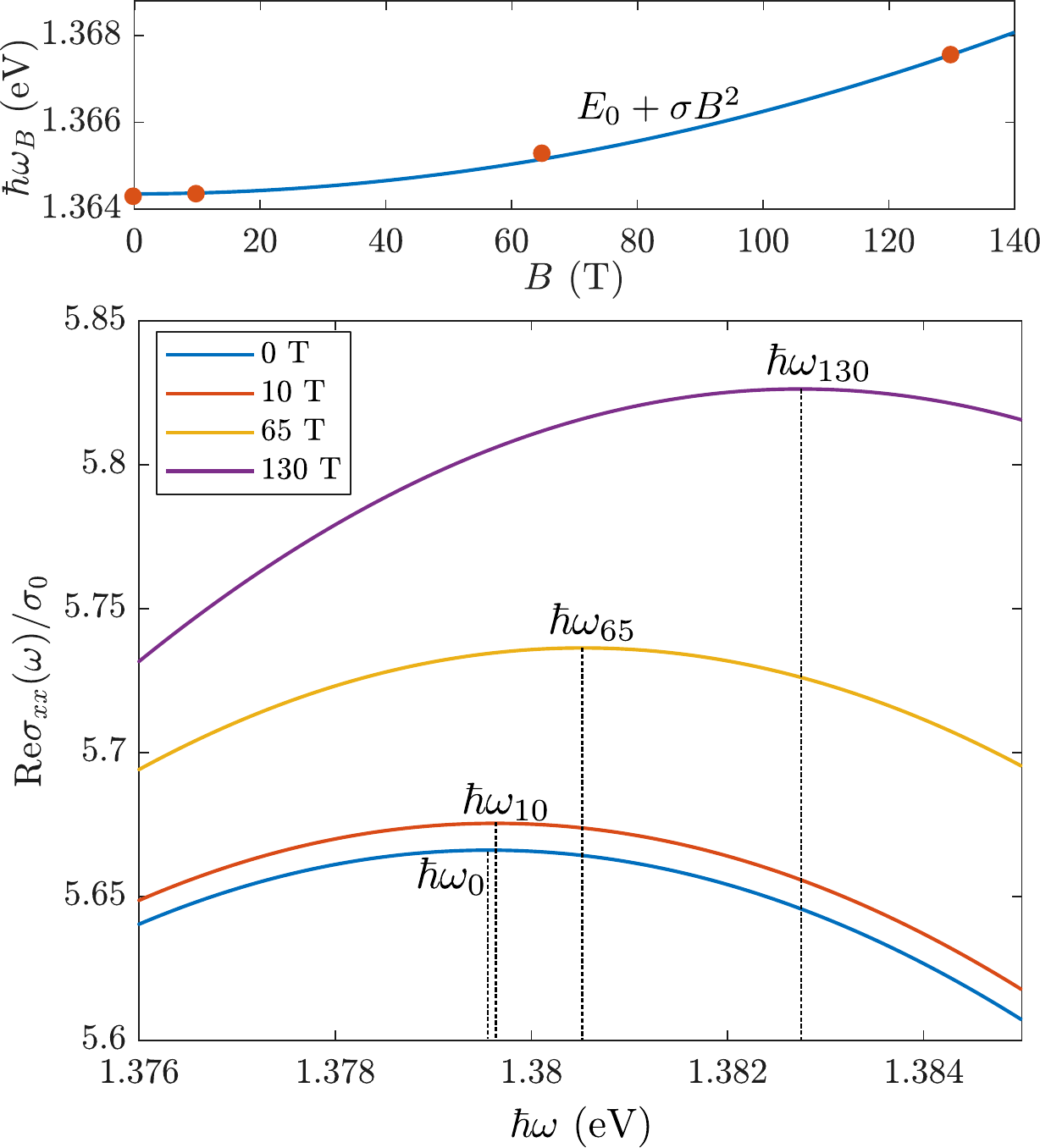}
  \caption{Diamagnetic shift of the peak associated with the $1s$ state of the $A$ exciton in WSe\textsubscript{2}. The vertical dashed lines indicate the peak position at different magnetic field strengths. The upper panel illustrates the fit of the function $E_0 + \sigma B^2$ (blue line) to the peak positions marked by the red dots.}\label{fig:exc-dia}
\end{figure}

\begin{table}[b]
  \centering
  \begin{tabular}{|l|c|c|c|}\hline
   $\kappa$ & $\sigma$ - BSE & $\sigma$ - Wannier & $\sigma$ - Exper.\\ \hline\hline
   1.00 & 0.22 & 0.13 & \\ \hline 
   1.55 & 0.24 & 0.15 & 0.18\cite{stier2016probing} \\ \hline 
   2.25 & 0.27 & 0.17 & 0.25\cite{stier2016probing} \\ \hline 
   3.30 & 0.31 & 0.19 & 0.32\cite{stier2016probing} \\ \hline 
   4.50 & 0.36 & 0.23 & 0.24\cite{liu2019magneto}, 0.31\cite {stier2018magnetooptics}\\ \hline\hline 
\end{tabular}
\caption{Calculated and experimental values of $\sigma$ in units of $\mu$eV/T\textsuperscript{2} for the $1s$ state of the $A$ exciton in WSe\textsubscript{2}. The first column is the values calculated using the approach presented in this paper, while the second column is values calculated using the Wannier model from Ref. \onlinecite{have2019monolayer}.}\label{tab:dia-shift}
\end{table}

In the low-field limit, the magnetic field dependence of the energy of $s$-type excitons can be described by the relation $E_B \approx E_0 + \sigma B^2$, where $E_0$ is the unperturbed exciton energy and $\sigma$ is the diamagnetic shift coefficient. The quadratic diamagnetic shift of the exciton peaks is illustrated in Fig.~\ref{fig:exc-dia} for the $A$ exciton in WSe\textsubscript{2}. This coincides precisely with the small shift observed in the diagonal conductivities in Fig.~\ref{fig:exc-mag}. As mentioned, the value of $\sigma$ is related to the spatial extent of the exciton. The relation is given by $\sigma = e^2\langle r^2\rangle/8\mu$, where $\sqrt{\langle r^2\rangle}$ is the root-mean-square (rms) radius of the exciton and $\mu$ is the reduced exciton mass. If $\mu$ is known, this relation allows for an experimental estimate of the exciton size. As the Lanczos method only provides the optical conductivity, and not the exciton energies, we compute the shift coefficient by following the exciton peak in the spectra as the field strength changes. The shift of the exciton peak is then fitted to a parabola, and the diamagnetic shift coefficient is found. Doing this for the $A$ exciton peak of freestanding WSe\textsubscript{2}, we find a $\sigma$-value of 0.22 $\mu$eV/T\textsuperscript{2}. Using the same effective masses as applied to find the TB parameters, we find an rms radius of 1.52 nm for the $A$ exciton.

The dielectric environment is expected to affect the size of the diamagnetic shift. Increasing $\kappa$ results in less tightly bound excitons, thus having a larger radius. This consequently results in larger diamagnetic shift coefficients. This effect was studied experimentally in Ref.~\onlinecite{stier2016probing}. In Table~\ref{tab:dia-shift}, we summarize our findings with regard to the effect of the dielectric environment on the diamagnetic shift coefficient. We have also included values computed from the Wannier model presented in Ref.~\onlinecite{have2019monolayer}. The Wannier model consistently underestimates $\sigma$, when comparing to the experimental values and the values computed using the nanoribbon approach. The explanation for this is found in the fact that the Bloch overlaps are disregarded in the Wannier model. This causes the excitons to be stronger bound in the Wannier framework than in BSE framework and, consequently, have smaller diamagnetic shift coefficients. This shows the importance of including the Bloch overlaps when modeling magnetoexcitons. Comparing the diamagnetic coefficients calculated using the nanoribbon approach to the experimental results, we observe a better agreement. 

\section{Summary}
\label{sec:summary}
In summary, we have used nanoribbons as a theoretical tool for the study of the magneto-optical response of monolayer TMDs. We have shown that by increasing the width of the nanoribbons the optical response will converge to that of a 2D monolayer. This has proven to be useful for including excitonic effects in the calculation of the magneto-optical response of TMDs, since a strict 2D calculation is not currently feasible. Beginning from a simple tight-binding model, we added excitonic effects in the framework provided by the Bethe-Salpeter equation. The linear optical conductivity was calculated effectively using the Lanczos-Haydock routine. We found that a 15-16 nm wide nanoribbon system is sufficient for a reasonable convergence of the optical response.

Using this approach, we are able to compute the excitonic Hall conductivity of monolayer TMDs. The calculated Hall conductivity spectra can be used to compute Faraday rotation in monolayer TMDs, an important magneto-optical effect. We also evaluated the diamagnetic shift coefficient, which provides a useful quantity for evaluating the size of excitons. So far, the experimentally determined diamagnetic shift coefficients have only been compared to theoretical results based on effective mass models. But our approach provide the option of going beyond effective mass models when analyzing experimental data. We compared the theoretical diamagnetic shift coefficient given by our calculation to values calculated using a Wannier model and to recent experimentally determined coefficients. The comparison with the values computed from the Wannier model showed the importance of including Bloch overlaps, while the comparison with experimental values showed a very good agreement between our calculations and the experimental results.

Finally, another potential use of the approach presented in this paper is as a benchmark for future strict 2D models. As it is currently not possible to compute the excitonic Hall conductivities in any 2D model, the Hall conductivity presented here can provide a reference when attempting to develop new models.

\section*{Acknowledgments}
J.H. and T.G.P gratefully acknowledge financial support by the QUSCOPE Center, sponsored by the Villum Foundation. Additionally, T.G.P. is supported by the Center for Nanostructured Graphene (CNG), which is sponsored by the Danish National Research Foundation, Project No. DNRF103. N.M.R.P. acknowledges support from the European Commission through the project “Graphene-Driven Revolutions in ICT and Beyond” (Ref. No. 785219), COMPETE2020, PORTUGAL2020, FEDER and the Portuguese Foundation for Science and Technology (FCT) through project POCI- 01-0145-FEDER-028114 and in the framework of the Strategic Financing UID/FIS/04650/2013.


\bibliographystyle{aipnum4-1}
\bibliography{references}

\appendix
\begin{widetext}
\section{Electron-hole interaction matrix elements for nanoribbons}\label{app:commutators}
In this appendix, we will find an expression for the matrix-elements in Eq.~\eqref{eq:eh-mel} for the nanoribbon geometry. We begin by considering the product of two tight-binding states, such as the ones in Eq.~\eqref{eq:eh-mel}. Exploiting the fact that the atomic orbitals are localized and orthogonal, we can write 
\begin{equation}
  \phi_{\alpha ks}^{*}(\mathbf{r})\phi_{\beta k's}(\mathbf{r}) \approx \frac{1}{N_{uc}}\sum_{n, X} I^{n,s}_{\alpha k, \beta k} e^{i(k'-k)X}\varphi^2_n(\mathbf{r}-X\hat{x}),
\end{equation}
where $N_{uc}$ is the number of unit cells, $X$ is the location of the unit cell in the periodic direction, $I^{n,s}_{\alpha k, \beta k} = C^{n *}_{\alpha ks}C^{n}_{\beta k's}$ are the products of the tight-binding eigenvector elements belonging to the $n$'th atomic orbital and $\varphi_n$ are the atomic orbitals. The $X$ sum runs over the location of the unit cells in the periodic direction. To evaluate the matrix-elements, we need integrals of the form
\begin{equation}
  U_{n,m}(X,X') =\iint \varphi^2_n(\mathbf{r}-X\hat{x})U(\mathbf{r}-\mathbf{r}')\varphi^2_m(\mathbf{r'}-X'\hat{x})\mathrm{d}^2\mathbf{r}\mathrm{d}^2\mathbf{r}'.
\end{equation}
For strongly localized atomic orbitals, we can assume the effective interaction
\begin{equation}
  U_{n,m}(X,X') \approx U_{n,m}^{\mbox{eff}}(X-X') \equiv -\frac{e^2}{8\varepsilon_0r_0}\left[H_0\left(\frac{\kappa \sqrt{(X-X')^2+Y_{nm}^2}}{r_0}\right)-Y_0\left(\frac{\kappa \sqrt{(X-X')^2+Y_{nm}^2}}{r_0}\right)\right].
\end{equation}
Here, $Y_{nm}$ denotes the difference in $y$-coordinates of the atomic site belonging to orbitals $n$ and $m$. This effective interaction is validated by its ability to recover the 2D results, as shown in the paper. In the following, it is advantageous to rewrite $U^{\mbox{eff}}_{n,m}(X-X')$ using an integral form of the Keldysh potential \cite{cudazzo2011dielectric}. This gives
\begin{equation}
  U^{\mbox{eff}}_{n,m}(X-X') = -\frac{e^2}{4\pi\varepsilon_0}\int_{0}^\infty\frac{1}{\sqrt{(r_0z)^2+(X-X')^2+Y_{nm}^2}}e^{-z\kappa}\mathrm{d}z.
\end{equation}
The interaction matrix-elements in Eq.~\eqref{eq:eh-mel} can then be approximated by
\begin{equation}
  W_{cvk,c'v'k'}^{s,s'} \approx \delta_{s,s'}\sum_{n,m}I^{n,s}_{c k, c' k'}I^{m,s}_{v' k',v k}\frac{1}{L}\int e^{i(k'-k)X}U^{\mbox{eff}}_{n,m}(X)\mathrm{d}X,\label{eq:eh-mel2}
\end{equation}
where we have converted the sum over $X$ to an integral and $L$ denotes the length of the system. Finally, we have to do the $X$ integration, which corresponds to taking the Fourier transform of the effective interaction. This gives
\begin{equation}
  \frac{1}{L}\int e^{i(k'-k)X}U^{\mbox{eff}}_{n,m}(X)\mathrm{d}X = -\frac{e^2}{2\pi L \varepsilon_0}\int_0^\infty\mathrm{d}z K_0\left(\sqrt{r_0^2z^2+Y_{nm}^2}\left|k-k'\right|\right)e^{-\kappa z},\label{eq:fourier}
\end{equation}
where $K_0$ denotes a modified Bessel function of the second kind. The remaining integral over $z$ can be evaluated numerically. Inserting Eq.~\eqref{eq:fourier} into Eq.~\eqref{eq:eh-mel2}, we obtain an expression for the interaction matrix-elements in the nanoribbon geometry.
\end{widetext}
\end{document}